\documentclass[preprint]{emulateapj-rtx4}

\shorttitle{Recently quenched galaxies at $z = 0.2 - 4.8$  in the COSMOS UltraVISTA field}
\shortauthors{Ichikawa and Matsuoka}

\begin{document}
\title{ recently quenched galaxies at {\scriptsize $z$} $= 0.2 - 4.8$  in the COSMOS UltraVISTA field}

\author{Akie Ichikawa\altaffilmark{1,2}}
\author{Yoshiki Matsuoka\altaffilmark{1,3}}

\altaffiltext{1}{Research Center for Space and Cosmic Evolution, Ehime University, 2-5 Bunkyo-cho, Matsuyama 790-8577, Japan.}
\altaffiltext{2}{Graduate School of Science \& Engineering, Ehime University, 2-5 Bunkyo-cho, Matsuyama 790-8577, Japan.}
\altaffiltext{3}{National Astronomical Observatory of Japan, 2-21-1 Osawa, Mitaka 181-8588, Japan.}

\email{ichikawa@cosmos.phys.sci.ehime-u.ac.jp}







\begin{abstract}

We present a new analysis of 
the stellar mass function and morphology of 
recently-quenched galaxies (RQGs), whose star formation has been recently quenched for some reason.
The COSMOS2015 catalog was exploited to select those galaxies 
at $0.2 < z < 4.8$, 
over 1.5 deg$^2$ of the Cosmic Evolution Survey (COSMOS) UltraVISTA field.
This is the first time that 
RQGs are consistently selected and studied 
in such a wide range of redshift.
We find increasing number density of RQGs with time in a broad mass range at $z>1$, while low-mass RQGs
start to grow very rapidly at $z < 1$.
We also demonstrate that the migration of RQGs may largely drive the evolution of the stellar mass function of passive galaxies.
Moreover, we find that the morphological type distribution
of RQGs are intermediate between those of star-forming and passive galaxies.
These results indicate that RQGs represent a major 
transitional phase of galaxy evolution, in which star-forming galaxies turn into passive galaxies, accompanied by 
the build up of spheroidal component.

\end{abstract}




\section{INTRODUCTION}
Galaxies are an important constituent of the universe, and are known to evolve throughout the cosmic time.
As such, it is crucial to disentangle the process of their formation and evolution, in order to better understand the history of the universe.
There are two major classes of galaxies, namely, star-forming and passive galaxies, but an evolutionary link between these two classes 
is still not well understand.
A seminal work by \cite{2003MNRAS.341...54K} demonstrated that the two classes have distinct stellar mass distributions, with the star-forming 
class dominating low-mass galaxies (stellar mass $M \la 10^{10.5} M_\odot$).
Transition from star-forming to passive galaxies must be accompanied by quenching of star formation in any form, whose process(es) may depend on
the properties of individual galaxies, such as stellar mass and surrounding environment 
\citep[e.g.,][]{2010ApJ...721..193P}. 

A potentially important population to probe the above quenching process(es) is post-starburst galaxies, characterized by the spectra being dominated by
A-type stars (\citealt{1983ApJ...270....7D}).
These galaxies are thought to be in a short period along the galaxy evolution, 
when the precedent active star formation was recently quenched . 
Their lifetime is estimated to be 0.1 - 0.6 Gyr \citep{2009MNRAS.395..144W, 2011ApJ...741...77S}.
%
Post-starburst galaxies are frequently observed with spectral signatures of active galactic nuclei (AGNs), which have been found in 
X-ray, optical, and/or mid-infrared (IR) wavelengths
\citep{2009ApJ...703..150B, 2016arXiv160800256A}.
This may suggest that
the star-formation quenching may be driven, at least partly, by energy input from AGNs \citep[see also][]{2014ApJ...780..162M, 2015ApJ...811...91M}.
\cite{2011ApJ...741..106C} found that post-starburst quasars are split into an equal number of spiral and early-type host galaxies, and that the 
luminous objects tend to exhibit morphological disturbance, which may suggest the merger origin of these systems.
High fractions of merger signatures were also reported for post-starburst galaxies by, e.g., \citet{1996ApJ...466..104Z}, \citet{2008ApJ...688..945Y}, 
and \citet{2012MNRAS.420.1684W}.
On the other hand, the majority of post-starburst galaxies are found in field environment \citep{1996ApJ...466..104Z}, and
no clear dependence of morphological evolution on the surrounding environment has been found for this class of galaxies 
(e.g., \citealt{2015ApJ...798...52V}).

Meanwhile, investigation of post-starburst galaxies is still limited compared to those of star-forming and passive galaxies. 
This is partly due to a much smaller number density of post-starburst galaxies than the other two classes; roughly 0.01 -- 1 \% of local galaxies are found
in post-starburst phase, depending on the selection criteria \citep[e.g.,][]{1996ApJ...466..104Z, 2004ApJ...602..190Q, goto07}.
This problem may be alleviated by selecting 
post-starburst analogs with photometry data alone, which enables to construct a much larger sample of galaxies
than would be possible with secure spectroscopic data.
Indeed, such photometrically-selected ``recently-quenched galaxies" (RQGs) have been studied at $z < 2$, with the imaging data from, e.g.,
the NEWFIRM Medium-Band Survey \citep{whitaker11, 2012ApJ...745..179W} and the UKIDSS \citep[][]{lawrence07} Deep Survey 
\citep{2016MNRAS.463..832W}.

In the present work, we exploit the exquisite multi-band photometry data from the Cosmic Evolution Survey (COSMOS) in order to study RQGs. 
The COSMOS field has been observed to very deep depths over $\sim$2 deg$^2$, across the electromagnetic spectrum from X-ray to radio 
wavelengths \citep{2007ApJS..172....1S}. 
In particular, deep near-IR data (see below) enable us to construct a robust sample of RQGs out to very high redshift; in this study,
we investigate the stellar mass function and morphology of RQGs at $0.2 < z < 4.8$.
We assume the cosmological parameters of $\Omega_m = 0.3$, $\Omega_\Lambda = 0.7$, and $H_0 = 70 $ km s$^{-1}$ Mpc$^{-1}$ throughout this work.
All magnitudes are presented in the AB system \citep{oke83}, unless otherwise noted.


\section{sample selection} 

This work exploits the COSMOS2015 catalog  (\citealt{2016ApJS..224...24L}), in which the source detection is based on the near-IR data taken from the
UltraVISTA survey (\citealt{2012A&A...544A.156M}) data release (DR) 2.
The UltraVISTA covers 1.5~deg$^2$ of the COSMOS field, with the DR2 limiting magnitudes (3$\sigma$ depth in an aperture of 2\arcsec diameter) of
$\sim$25 mag in the $YJHK_{\rm S}$ bands.
The effective area are 0.46 and 0.92 deg$^2$ in the Ultra-deep  (UD) and Deep fields, respectively, 
after removing masked regions around bright sources. 
Photometric redshifts of galaxies were measured with
$Le Phare$ (\citealt{2002MNRAS.329..355A}; \citealt{2006A&A...457..841I}), with the spectral templates taken from 
\cite{2007ApJ...663...81P} and \citet[][hereafter BC03]{2003MNRAS.344.1000B}.
These templates do not explicitly include a spectrum of post-starburst galaxies \citep[such as that from][]{2010ApJ...722L..64K}, but post-starburst 
galaxies may be accounted for partly by the above BC03 templates, with the assumed starburst ages ranging from 0.03 to 3 Gyr.
Here we simply assume that the redshifts of our RQGs are reasonably determined with the adopted templates, and defer a more detailed analysis of photometric-redshift
accuracy to a future work.
The present work uses photometric redshifts for all the galaxies, and does not exploit spectroscopic information of any individual galaxies. 
Absolute magnitudes were derived with the best-fit spectral templates, while stellar masses were estimated with the BC03 models, assuming the
\citet{chabrier03} initial mass function.
We removed X-ray sources in this work, given that their redshift, absolute magnitude, and stellar mass estimates are relatively uncertain.




The samples of galaxies used in this work are selected as follows.
First, we extract all the galaxies
from the COSMOS2015 catalog in the UltraVISTA fields, 
excluding those in masked regions or with matched X-ray detection. 
We then remove objects with  
absolute magnitudes $NUV > -25$ mag,
since a small fraction of COSMOS2015 objects have catastrophically bright $NUV$ magnitudes for unknown reason (C. Laigle, private communication).
This work makes use of the limiting masses of 90\% completeness ($M^{\rm full}_{\rm lim}$) presented in \citet{2016ApJS..224...24L},
and selects galaxies more massive than 
$M^{\rm full}_{\rm lim}$
at each redshift.
A discussion about possible selection effects is found in the following section.

Next, we classify the extracted galaxies into star-forming, passive, and RQG classes based on their colors.
Broad-band selection of these classes of galaxies have been commonly performed
(e.g., \citealt{2009A&A...501...15F}; \citealt{2009ApJ...707.1595D}; \citealt{2010ApJ...709..644I}).
In particular, 
the rest-frame $U$, $V$, and $J$-band colors have been 
 frequently used for galaxy classification 
 (e.g., \citealt{2015ApJ...804L...4M}). 
The spectral energy distributions (SEDs) 
of passive galaxies have a strong Balmer/4000~$\rm{\AA}$ break,
which effectively separates star-forming and passive galaxies in a broad-band color space.
In this work, we adopt the color selection with the rest-frame $NUV$, $r$, and $J$ bands (\citealt{2013A&A...556A..55I}). 
The COSMOS2015 $U$ band 
refers to the 
CFHT/MegaPrime $u^*$ band, whose filter transmission
extends to a slightly longer wavelength than the 4000~$\rm{\AA}$ break;
this makes the $NUV$ band a better tracer of 
the blue side of the 4000~$\rm{\AA}$ break, in a given spectrum.

\begin{figure}[]
\begin{center}
\includegraphics
[width=80mm,angle=0]
{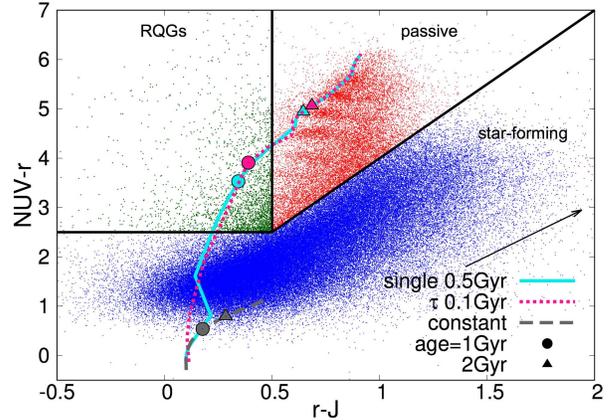}
\end{center}
\caption{Rest-frame $NUV$-$r$-$J$ color diagram. The blue, red and green dots represent 
star-forming galaxies, passive galaxies, and RQGs, respectively.
The cyan, pink, and gray 
lines represent the
color tracks of simple stellar populations by the single 
burst model, exponentially declining
model, and constant star formation model, respectively (see text).
The big dots and triangles mark the stellar ages of  
1 and 2 Gyr, respectively, in
each model.
The arrow represents
dust extinction corresponding to $A_{\rm V} = 1$ mag (\citealt{2000ApJ...533..682C}). 
}
\label{2color}
\end{figure}

In order to establish the color selection criteria, 
we calculated the typical colors of galaxies with the BC03 models.
We assumed simple stellar populations with three different star-formation histories, i.e., a single starburst with 0.5~Gyr duration, an exponentially declining 
star formation rate ($\propto e^{-\tau}$) with $\tau$ = 0.1~Gyr, and a constant star formation rate. 
The results of these calculates are presented in Figure~\ref{2color}.
The SEDs of RQGs have a strong Balmer/4000 $\rm{\AA}$ break
contributed by A-type stars, which 
makes their $NUV-r$ colors  
redder than those of star-forming galaxies
(e.g. \citealt{2007ApJS..173..342M}, \citealt{2013A&A...556A..55I}).
In addition, this diagram allows to 
distinguish RQGs and passive galaxies,  
due to the bluer SEDs (bluer $r-J$ colors) of RQGs
at $>$4000 \AA\ 
compared to passive galaxies.

Figure~\ref{2color} demonstrates that RQGs, 
which corresponds to the age around 1~Gyr in the single burst or the exponentially declining model,
can be selected at the top left corner of the diagram.
Our selection criteria of star-forming galaxies, passive galaxies, and RQGs are represented by the black solid lines, which are defined as:

\begin{eqnarray}
(NUV-r)<2.5  {\rm ~or~} (NUV-r)<3(r-J)+1 \nonumber
\end{eqnarray}
for star-forming galaxies,

\begin{eqnarray}
(NUV-r)>2.5  &{\rm ~and~}& (NUV-r)> 3(r-J)+1 \nonumber \\ 
&{\rm ~and~}& (r-J)>0.5 \nonumber
\end{eqnarray}
for passive galaxies, and

\begin{eqnarray}
(NUV-r)>2.5  &{\rm ~and~}& (NUV-r)>3(r-J)+1 \nonumber \\ 
&{\rm ~and~}& (r-J)<0.5 \nonumber
\end{eqnarray}
for RQGs. 
The separation between star-forming and passive galaxies (the diagonal line in Figure~\ref{2color}) is identical to that used by 
\cite{2013A&A...556A..55I}, which is nearly 
parallel to the direction of dust reddening. 
The separation between star-forming and RQGs (the horizontal line in Figure~\ref{2color}) corresponds to the epoch of 0.2 Gyr after the star formation is 
stopped in the single burst model.
This is based on the hydrodynamic simulations of galaxy mergers presented by \citet{2011ApJ...741...77S}, who demonstrated that
a galaxy may be selected as a post-starburst system after $\sim$0.2 Gyr of the merger (and starburst) event.

\begin{figure*}[]
\begin{center}
\includegraphics
[width=160mm,angle=0]
{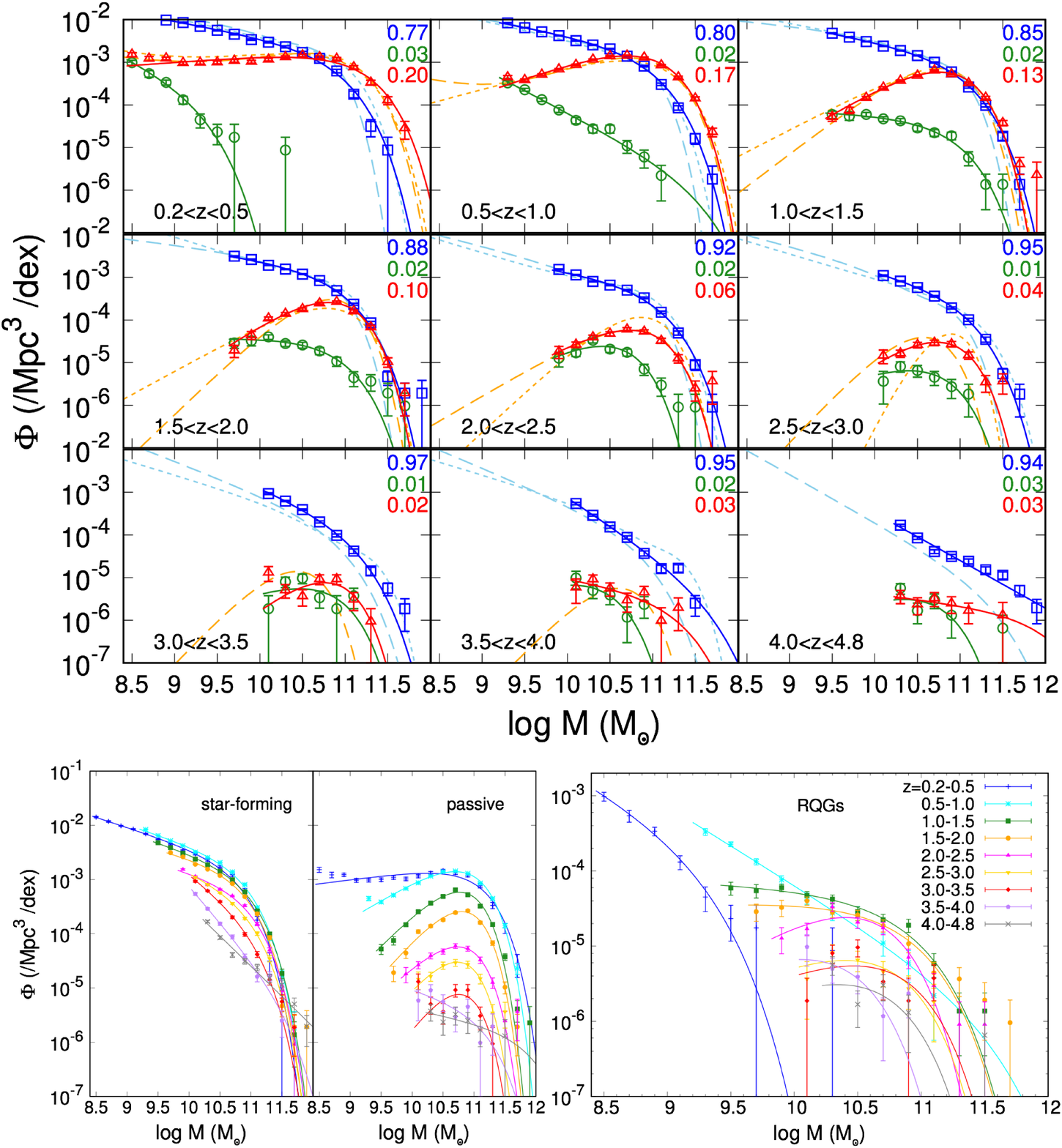}
\end{center}
\caption{
Top nine panels: the stellar mass functions derived in this work, as well as in the previous studies in close (but not exactly the same) redshift ranges.
The blue, red, and green points and lines represent the star-forming galaxies, passive galaxies, and RQGs in this work, respectively.
The cyan and orange lines represent those of star-forming and passive galaxies in the previous studies, 
taken from \cite{2017arXiv170102734D} (long dashed lines) and \cite{2013A&A...556A..55I} (short dashed lines). 
The number fractions of the three classes of galaxies, integrated over the whole measured mass range, are indicated 
at the top right of each panel (the color coding is the same as for the lines). 
Bottom three panels: the stellar mass functions 
of the star-forming galaxies (left), passive galaxies (middle), and RQGs (right) at different redshifts. 
The lines represent the best-fit Shechter functions, restricted to the mass range above the 90\% completeness mass at each redshift.
We have no ability to probe the lower-mass regime than plotted here, which would require deeper near-IR observations than available now.
}
\label{compare_presious}
\end{figure*}

\begin{table*}[t]
\begin{center}
\caption{
Best-fit parameters of the Schechter 
stellar mass functions 
}
\begin{tabular}{c c@{$\pm$}l c@{$\pm$}l c@{$\pm$}l c c }%
\tablewidth{0pt}%
\hline%
\hline%
    \multicolumn{8}{c}{star-forming galaxies}\\  
redshift & \multicolumn{2}{c}{ {\it $\Phi^*$} ($10^{-3}$/${\rm Mpc}^3$/dex)} & \multicolumn{2}{c}{ log$M^* (M_{\odot})$} & \multicolumn{2}{c}{ $\alpha$} & $\chi^2_{reduced}$ & Number\footnote{Number of galaxies in each redshift bin.}
 \\  \hline
$0.2 \leq z < 0.5$	&	0.860	&	0.070	&	10.81	&	0.03	&	-1.37	&	0.01	&	2.14 & 9750 \\
$0.5 \leq z < 1.0$ 	&	0.888	&	0.090	&	10.89	&	0.03	&	-1.37	&	0.03	&	6.65 & 23843 \\
$1.0 \leq z < 1.5$ 	&	0.605	&	0.036	&	10.93	&	0.02	&	-1.38	&	0.02	&	1.81 & 17956 \\
$1.5 \leq z < 2.0$ 	&	0.620	&	0.069	&	10.86	&	0.03	&	-1.30	&	0.04	&	5.93 & 12902 \\
$2.0 \leq z < 2.5$ 	&	0.302	&	0.055	&	10.94	&	0.05	&	-1.32	&	0.07	&	5.92 & 6846 \\
$2.5 \leq z < 3.0$ 	&	0.212	&	0.022	&	10.94	&	0.03	&	-1.50	&	0.04	&	0.67 & 4050 \\
$3.0 \leq z < 3.5$ 	&	0.088	&	0.026	&	10.99	&	0.07	&	-1.83	&	0.08	&	1.34 & 2694 \\
$3.5 \leq z < 4.0$ 	&	0.003	&	0.005	&	11.53	&	0.32	&	-2.30	&	0.12	&	1.74 & 1213 \\
$4.0 \leq z < 4.8$ 	&	0.000	&	0.001	&	13.71	&	23.79	&	-2.11	&	0.14	&	3.37 & 896\\ \hline \hline
    \multicolumn{8}{c}{recently-quenched galaxies (RQGs)}\\  
redshift & \multicolumn{2}{c}{ {\it $\Phi^*$} ($10^{-3}$/${\rm Mpc}^3$/dex)} & \multicolumn{2}{c}{ log$M^* (M_{\odot})$} & \multicolumn{2}{c}{ $\alpha$} & $\chi^2_{reduced}$ & Number$\rm{^a}$ \\  \hline
$0.2 \leq z < 0.5$	&	1.373	&	1.275	&	9.15	&	0.20	&	-1.92	&	0.30	&	0.68 & 185 \\
$0.5 \leq z < 1.0$ 	&	0.009	&	0.014	&	11.45	&	0.53	&	-2.04	&	0.07	&	0.72 & 549 \\
$1.0 \leq z < 1.5$ 	&	0.222	&	0.041	&	10.80	&	0.07	&	-1.09	&	0.08	&	0.62 & 356 \\
$1.5 \leq z < 2.0$ 	&	0.194	&	0.054	&	10.77	&	0.11	&	-0.94	&	0.16	&	0.99 & 249 \\
$2.0 \leq z < 2.5$ 	&	0.283	&	0.030	&	10.37	&	0.11	&	0.08	&	0.39	&	1.47 & 185 \\
$2.5 \leq z < 3.0$ 	&	0.073	&	0.020	&	10.53	&	0.23	&	-0.26	&	0.64	&	0.60 & 44 \\
$3.0 \leq z < 3.5$ 	&	0.062	&	0.019	&	10.60	&	0.21	&	 \multicolumn{2}{c}{-0.26}		&	1.54 & 44 \\
$3.5 \leq z < 4.0$ 	&	0.075	&	0.030	&	10.18	&	0.13	&	 \multicolumn{2}{c}{-0.26}		&	1.09 & 25 \\
$4.0 \leq z < 4.8$ 	&	0.035	&	0.015	&	10.47	&	0.27	&	 \multicolumn{2}{c}{-0.26}		&	1.45 & 27 \\ \hline \hline
    \multicolumn{8}{c}{passive galaxies}\\  
redshift & \multicolumn{2}{c}{ {\it $\Phi^*$} ($10^{-3}$/${\rm Mpc}^3$/dex)} & \multicolumn{2}{c}{ log$M^* (M_{\odot})$} & \multicolumn{2}{c}{ $\alpha$} & $\chi^2_{reduced}$ & Number$\rm{^a}$  \\  \hline
$0.2 \leq z < 0.5$	&	0.870	&	0.107	&	11.07	&	0.05	&	-0.85	&	0.04	&	5.04 & 2615 \\
$0.5 \leq z < 1.0$ 	&	1.565	&	0.057	&	10.86	&	0.02	&	-0.31	&	0.03	&	3.24 & 7686 \\
$1.0 \leq z < 1.5$ 	&	0.679	&	0.022	&	10.69	&	0.03	&	0.17	&	0.07	&	3.43 & 3949 \\
$1.5 \leq z < 2.0$ 	&	0.287	&	0.010	&	10.67	&	0.03	&	0.34	&	0.09	&	2.28 & 1888 \\
$2.0 \leq z < 2.5$ 	&	0.067	&	0.002	&	10.70	&	0.03	&	0.08	&	0.08	&	0.38 & 472 \\
$2.5 \leq z < 3.0$ 	&	0.033	&	0.003	&	10.58	&	0.06	&	0.43	&	0.26	&	0.83 & 219 \\
$3.0 \leq z < 3.5$ 	&	0.007	&	0.003	&	10.53	&	0.26	&	0.67	&	1.16	&	2.33 & 66 \\
$3.5 \leq z < 4.0$ 	&	0.002	&	0.003	&	11.16	&	0.78	&	-1.34	&	0.63	&	1.07 & 41 \\
$4.0 \leq z < 4.8$ 	&	0.001	&	0.001	&	11.91	&	1.44	&	-1.24	&	0.32	&	0.30 & 38 \\ \hline
 \end{tabular}
  \end{center}
\end{table*}

\section{RESULTS AND DISCUSSION}
We calculated
the number density of galaxies separately
for the UltraVISTA UD and Deep area, above the 90\% completeness mass, and combined them with inverse-variance weighting
if both the UD and Deep measurements are available in a given mass bin.
Then we fitted
the observed 
mass functions with the
Schechter function (\citealt{1976ApJ...203..557S}), which is defied as 
\begin{eqnarray}
 \Phi(M) {\rm d}M = {\rm ln(10)} \Phi^* [10^{(M-M^*)(\alpha + 1)} ] {\rm exp} [-10^{(M-M^*)}] {\rm d}M . \nonumber
\end{eqnarray}
Given the small sample size of RQGs at $z > 3$, we fixed $\alpha$ of those galaxies to the value obtained at $2.5 < z < 3.0$.

Figure \ref{compare_presious} presents
the derived stellar mass functions 
at $z=0.2-4.8$. 
The top nine panels compare the functions of the three classes of galaxies, while the bottom three panels compare those in different redshift bins.
The present stellar mass functions 
are broadly consistent with the past measurements for star-forming and passive galaxies (\citealt{2013A&A...556A..55I}; \citealt{2017arXiv170102734D}).
The best-fit parameters of the Schechter function  
are summarized in Table 1.

At the top right in each 
of the top panels, we report 
the number fraction of the three classes of galaxies,
integrated over the whole measured mass range.
The fraction of star-forming galaxies increases with time toward $z \sim 3$, which is close
to the peak epoch of the cosmic star-formation rate density (\citealt{2014ARA&A..52..415M}),
and then starts to decrease toward the present-day universe.
The fraction of passive galaxies increases with time across the redshift, reflecting
the decreasing fraction of star-forming galaxies. 
The fraction of 
RQGs is much smaller than those of the other two classes, 
and stays nearly constant at the level of $\sim$2~\%.

This is the first time that the stellar mass functions of
RQGs are consistently measured in such a wide range of redshift.
The mass functions at $z > 2$ have a clear peak at around log$(M/M_\odot) \sim10.5$, while those at $z < 1.5$ rise monotonically toward the low-mass end.
We found growing number density of RQGs with time at almost all the available masses at $z > 1$. 
The increase in number of log$(M/M_\odot) \sim 10.5$ galaxies decelerates 
at $1.0 < z  < 2.5$, while lower-mass galaxies continue to increase during the same period.
Then low-mass galaxies 
start to dominate the RQG population, resulting in the rapid steepening of the mass function toward the present-day universe. 
The deficit of massive RQGs in the local universe is consistent with the earlier finding by \cite{2016MNRAS.463..832W}.

Since RQGs are a short-lived population and are continuously formed by quenching starburst in star-forming galaxies, 
the above differential increase/decrease in number as a function of mass reflects
two factors.
One is the evolution of stellar mass function of star-forming galaxies, and the other is the efficiency of the quenching process(es) at each mass.
It is becoming clear that the star-formation quenching comes in (at least) two flavors.
The so-called mass quenching is efficient at log$(M/M_\odot) > 10.5$, while less massive galaxies are thought to be more sensitive to the environmental quenching \citep{2010ApJ...721..193P}. 
It would be interesting to see, with hydrodynamic simulations of quenching processes, whether the observed evolution of the stellar mass functions of 
star-forming galaxies and RQGs could be reproduced in a self-consistent way.
This is a subject of future works.

\begin{figure*}[]
\begin{center}
\includegraphics
[width=120mm,angle=270]
{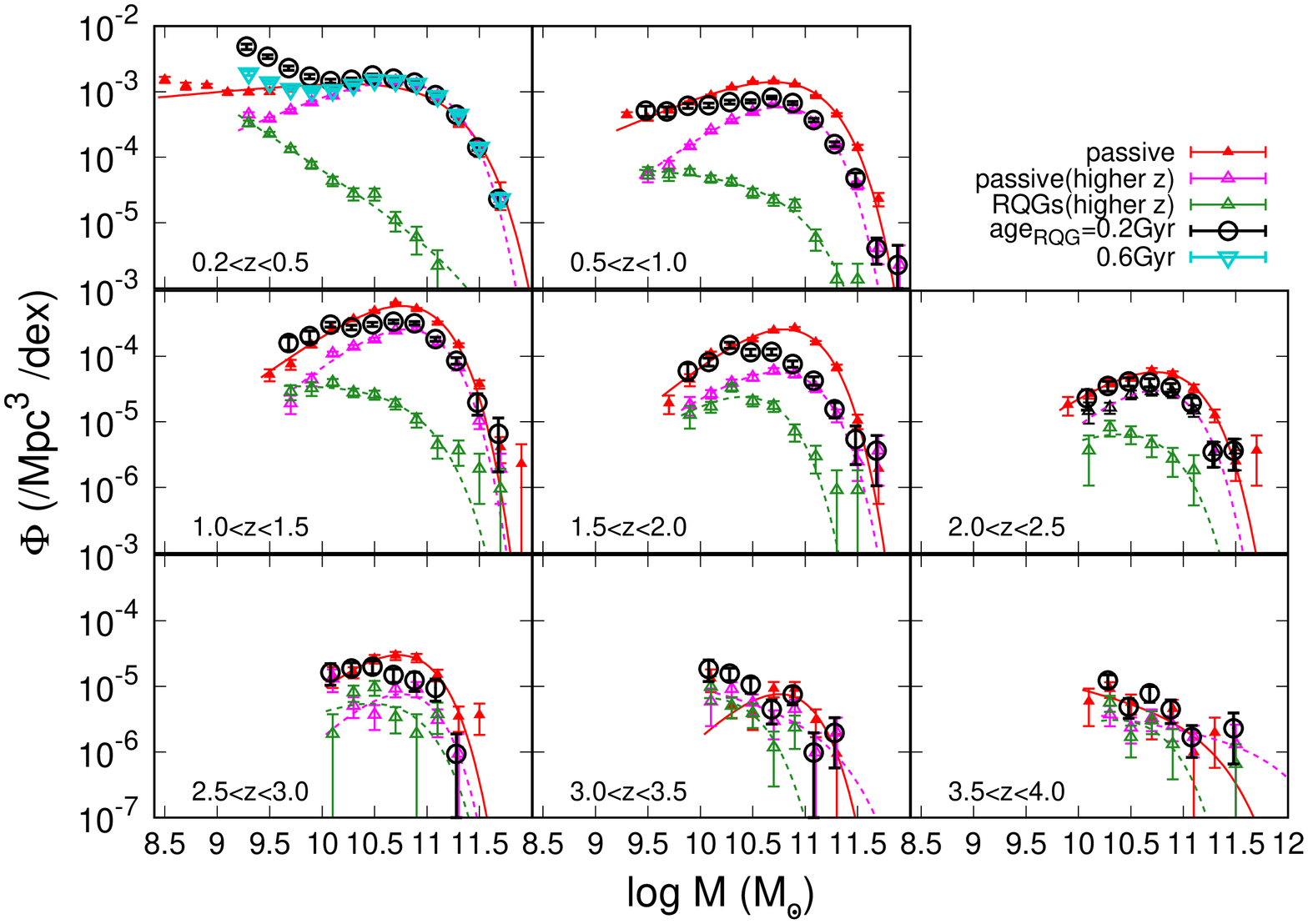}
\end{center}
\caption{
Observed stellar mass functions of passive galaxies (the red symbols and lines), compared to the simulated functions (black and cyan) by adding 
those of RQGs (green) to passive galaxies (pink) in the adjacent higher redshift bin.
The RQG lifetime is assumed to be 0.2~Gyr (black) or 0.6~Gyr (cyan).
}
\label{x-postSB}
\end{figure*}

The rapid increase of low-mass RQGs toward the local universe may be responsible for the observed flatterning of the stellar mass function of passive galaxies, 
at the low-mass end.
Following the recipe in \cite{2012ApJ...745..179W} and \cite{2016MNRAS.463..832W}, here we demonstrate that the evolution of the stellar mass functions of 
passive galaxies can indeed be reproduced, at least partly, by adding the observed RQGs to the passive galaxies that existed in the previous era
(i.e., in the adjacent higher redshift bin). 
Since RQGs are recognized as such only when the stellar population is
dominated by A-type stars, they have a relatively short lifetime.
We assume the typical RQG lifetime of 0.2~Gyr in this work, based on the recent estimate by \citet{2009MNRAS.395..144W, 2016MNRAS.463..832W}.
The results are presented in Figure~\ref{x-postSB}, which
suggests that most of new low-mass passive galaxies are formed by the migration of RQGs.
This is consistent with the earlier results by \cite{2016MNRAS.463..832W}.
However, we observe a clear deficit of new passive galaxies at high mass side.
This result may point to an additional path of passive galaxy evolution, e.g., dry mergers of galaxies.
\citet{2007ApJ...658...65C} pointed out an importance of dry mergers for later evolution of elliptical galaxies, after they were formed in a process similar to 
monolithic collapse at high redshift, in order to reconcile the scaling laws observed in the local universe.
On the other hand, the assumed RQG lifetime of 0.2~Gyr clearly overestimates the number of local passive galaxies at $z < 0.5$.
This is alleviated when we assume
the longer lifetime of 0.6~Gyr, which is the maximum value suggested by \citet{2009MNRAS.395..144W}.
Alternatively, this may indicate that a large fraction of RQGs are rejuvenated from passive galaxies with recent minor mergers, as suggested 
by \citet{2013ApJ...770...62D} and \citet{abramson13} for a sample of post-starburst galaxies at $z \le 0.5$.
Roughly a half of the observed RQGs must be formed by such a rejuvenation, in order to reproduce the evolution of passive galaxies
at $z < 0.5$ with the RQG lifetime of 0.2~Gyr.
Combined with the disappearance of massive RQGs in the local universe, the above results may indicate a fundamental difference in the formation and evolution of 
RQGs, at high and low redshifts.


Next we investigate the morphology of the three classes of galaxies.
We use the Zurich Structure \& Morphology Catalog v1.0, which
classifies structures of COSMOS galaxies based on the Zurich Estimator of Structural Types (ZEST) algorithm described in \citet{2007ApJS..172..406S}.
The ZEST classification has three main morphological classes,
namely, early, 
disk, and irregular classes.
In addition, bulgeness and irregularity\footnote{Bulgeness is defined with the S\'{e}rsic index, and is closely correlated with the bulge-to-disk ratio (B/D).
Irregularity is basically determined with visual inspection of surface flux density distribution, coupled with the principal component analysis in the ZEST algorithm.
See \citet{2007ApJS..172..406S} for more details.
} 
are measured for disk and elliptical galaxies, respectively. 
Here we concentrate on galaxies at $z < 2$, where the morphological classification is relatively complete and reliable, and with 
log$M_{*}\geq9.9$, i.e., above the 90\% mass completeness limit at $z = 2$.

\begin{figure*}[]
\begin{center}
\includegraphics
[width=180mm,angle=0]
{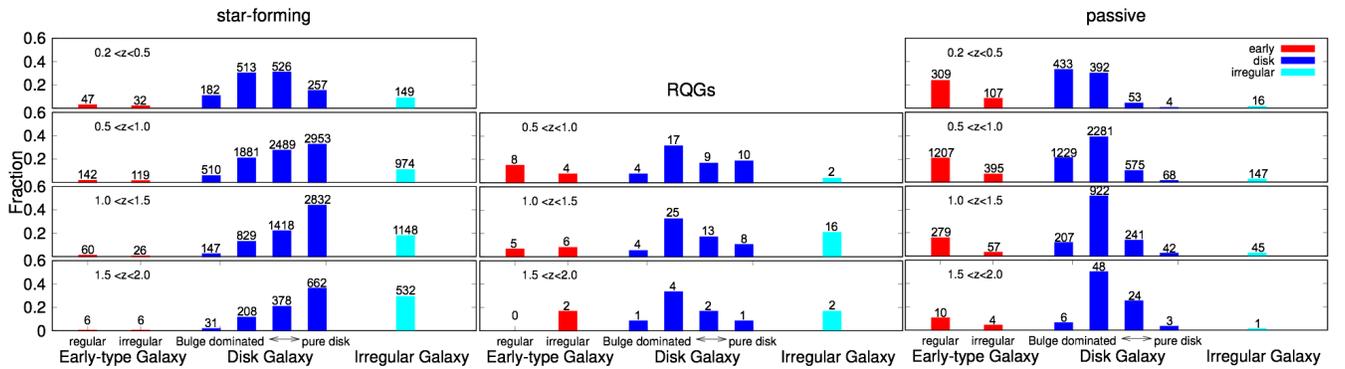}
\end{center}
\caption{
Fraction of each morphological type for
the star-forming galaxies (left panel), RQGs (middle panel), and passive galaxies (right panel), based on the ZEST classification.
The red, blue, and light-blue bars represent early, disk, and irregular-type galaxies, respectively, with subdivisions according to the irregularity and bulgeness (see text).
The number of galaxies included in each morphological type is indicated above each bar.
}
\label{MOAPHO}
\end{figure*}

The results of this morphological investigation are presented 
in Figures~\ref{MOAPHO}. As expected, 
star-forming galaxies are dominated by disk galaxies at all the redshifts.
Among the star-forming population, bulge-dominated disk galaxies and early-type galaxies increase their fractions with the passage of time, while
irregular galaxies decrease its fraction. 
Meanwhile, 
passive galaxies have a much larger fraction of 
early type morphology.
As in star-forming galaxies, there is a clear trend of pure disk galaxies transiting to bulge-dominated and early-type galaxies.
Overall, the results shown here
are consistent with the general picture, that  
disk galaxies dominate star-forming galaxies and passive galaxies have a higher fraction of early-type morphology.

On the other hand, the early- to disk-type ratios of RQGs are intermediate between those of the star-forming and passive galaxies.
No clear trend of morphological evolution is found for RQGs, except for the emergence of regular early type toward lower redshift.
This is consistent with the above argument, that RQGs are likely in a short transition phase from star-forming to passive galaxies.

Finally, we discuss the possible selection effects in the present study. 
Since our samples of galaxies are extracted from a near-IR selected, COSMOS2015 catalog, they are fairly complete down to the 90\% completeness masses,
presented in \citet{2016ApJS..224...24L} and used throughout this work.
Galaxies below the completeness masses are not considered, and so we cannot rule out the possibility that lower-mass galaxies behave quite differently
from those studied here. 
However, due to the exquisite depth of the UltraVISTA survey, the completeness masses are always below the characteristic masses ($M_*$) of the stellar mass
functions of the three classes of galaxies, out to the highest redshift considered in this work.
Hence the present results are fairly general, though deeper observations are necessary to probe the whole populations of RQGs and other galaxies,
down to very low-mass regime.



\section{SUMMARY}
We exploit the COSMOS2015 catalog to investigate the stellar mass function and morphology of recently quenched galaxies (RQGs) at $z = 0.2 - 4.8$, selected over 1.5 deg$^2$
of the COSMOS UltraVISTA survey field.
Our findings are summarized as follows.\\
1. Number density of RQGs increases with time in a broad mass range at $z > 1$. 
Then low-mass RQGs start to grow their number very rapidly toward the present-day universe, which may be due to a quenching process working effectively in low-mass
star-forming galaxies, such as the putative environmental quenching.\\
2. The evolution of the stellar mass function of passive galaxies may be driven largely by the migration of RQGs, assuming the RQG lifetime of 0.2 Gyr.
This migration naturally reproduces the flattening of the stellar mass function of passive galaxies over time.
However, additional evolutionary paths may be required for high-mass passive galaxies and for RQGs in the local universe, which might include dry mergers and
rejuvenation of passive galaxies.
 \\
3. The morphological early- to disk-type ratios of RQGs are intermediate between those of star-forming and passive galaxies. 
This is consistent with the argument that RQGs represent a short transitional phase of galaxy evolution, when star-forming galaxies turn into passive galaxies,
accompanied by the build up of spheroidal component. \\

\acknowledgments
We are grateful to the referee for very useful comments to improve this manuscript.
We thank Y. Taniguchi, M. Kajisawa and T. Nagao for inspiring us to carry out this research. 
Data analysis of this work was 
executed on the open use data analysis computer system at the Astronomy Data Center of the National Astronomical Observatory of Japan. 
YM was supported by JSPS KAKENHI Grant No. JP17H04830.


\begin{thebibliography}

\bibitem[Abramson et al.(2013)]{abramson13} Abramson, L.~E., Dressler, A., Gladders, M.~D., et al.\ 2013, \apj, 777, 124 
\bibitem[Alatalo et al.(2016)]{2016arXiv160800256A} Alatalo, K., Bitsakis, T., Lanz, L., et al.\ 2016, arXiv:1608.00256 
\bibitem[Arnouts et al.(2002)]{2002MNRAS.329..355A} Arnouts, S., Moscardini, L., Vanzella, E., et al.\ 2002, \mnras, 329, 355 
\bibitem[Brown et al.(2009)]{2009ApJ...703..150B} Brown, M.~J.~I., Moustakas, J., Caldwell, N., et al.\ 2009, \apj, 703, 150 
\bibitem[Bruzual \& Charlot(2003)]{2003MNRAS.344.1000B} Bruzual, G., \& Charlot, S.\ 2003, \mnras, 344, 1000 
\bibitem[Cales et al.(2011)]{2011ApJ...741..106C} Cales, S.~L., Brotherton, M.~S., Shang, Z., et al.\ 2011, \apj, 741, 106
\bibitem[Calzetti et al.(2000)]{2000ApJ...533..682C} Calzetti, D., Armus, L., Bohlin, R.~C., et al.\ 2000, \apj, 533, 682 
\bibitem[Chabrier(2003)]{chabrier03} Chabrier, G.\ 2003, \pasp, 115, 763 
\bibitem[Ciotti et al.(2007)]{2007ApJ...658...65C} Ciotti, L., Lanzoni, B., \& Volonteri, M.\ 2007, \apj, 658, 65 
\bibitem[Davidzon et al.(2017)]{2017arXiv170102734D} Davidzon, I., Ilbert, O., Laigle, C., et al.\ 2017, arXiv:1701.02734 
\bibitem[Dressler \& Gunn(1983)]{1983ApJ...270....7D} Dressler, A., \& Gunn, J.~E.\ 1983, \apj, 270, 7 
\bibitem[Dressler et al.(2013)]{2013ApJ...770...62D} Dressler, A., Oemler, A., Jr., Poggianti, B.~M., et al.\ 2013, \apj, 770, 62
\bibitem[Drory et al.(2009)]{2009ApJ...707.1595D} Drory, N., Bundy, K., Leauthaud, A., et al.\ 2009, \apj, 707, 1595 
\bibitem[Fioc \& Rocca-Volmerange(1997)]{1997A&A...326..950F} Fioc, M., \& Rocca-Volmerange, B.\ 1997, \aap, 326, 950 
\bibitem[Fontana et al.(2009)]{2009A&A...501...15F} Fontana, A., Santini, P., Grazian, A., et al.\ 2009, \aap, 501, 15 
\bibitem[Goto(2007)]{goto07} Goto, T.\ 2007, \mnras, 381, 187 
\bibitem[Ilbert et al.(2006)]{2006A&A...457..841I} Ilbert, O., Arnouts, S., McCracken, H.~J., et al.\ 2006, \aap, 457, 841 
\bibitem[Ilbert et al.(2010)]{2010ApJ...709..644I} Ilbert, O., Salvato, M., Le Floc'h, E., et al.\ 2010, \apj, 709, 644 
\bibitem[Ilbert et al.(2013)]{2013A&A...556A..55I} Ilbert, O., McCracken, H.~J., Le F{\`e}vre, O., et al.\ 2013, \aap, 556, A55 
\bibitem[Kauffmann et al.(2003)]{2003MNRAS.341...54K} Kauffmann, G., Heckman, T.~M., White, S.~D.~M., et al.\ 2003, \mnras, 341, 54 
\bibitem[Kriek et al.(2010)]{2010ApJ...722L..64K} Kriek, M., Labb{\'e}, I., Conroy, C., et al.\ 2010, \apjl, 722, L64 
\bibitem[Laigle et al.(2016)]{2016ApJS..224...24L} Laigle, C., McCracken, H.~J., Ilbert, O., et al.\ 2016, \apjs, 224, 24 
\bibitem[Lawrence et al.(2007)]{lawrence07} Lawrence, A., Warren, S.~J., Almaini, O., et al.\ 2007, \mnras, 379, 1599 
\bibitem[Madau \& Dickinson(2014)]{2014ARA&A..52..415M} Madau, P., \& Dickinson, M.\ 2014, \araa, 52, 415 
\bibitem[Martin et al.(2007)]{2007ApJS..173..342M} Martin, D.~C., Wyder, T.~K., Schiminovich, D., et al.\ 2007, \apjs, 173, 342 

\bibitem[Matsuoka et al.(2014)]{2014ApJ...780..162M} Matsuoka, Y., Strauss, M.~A., Price, T.~N., III, \& DiDonato, M.~S.\ 2014, \apj, 780, 162 
\bibitem[Matsuoka et al.(2015)]{2015ApJ...811...91M} Matsuoka, Y., Strauss, M.~A., Shen, Y., et al.\ 2015, \apj, 811, 91 
\bibitem[McCracken et al.(2012)]{2012A&A...544A.156M} McCracken, H.~J., Milvang-Jensen, B., Dunlop, J., et al.\ 2012, \aap, 544, A156 
\bibitem[Mendel et al.(2015)]{2015ApJ...804L...4M} Mendel, J.~T., Saglia, R.~P., Bender, R., et al.\ 2015, \apjl, 804, L4 
\bibitem[Oke \& Gunn(1983)]{oke83} Oke, J.~B., \& Gunn, J.~E.\ 1983, \apj, 266, 713 
\bibitem[Peng et al.(2010)]{2010ApJ...721..193P} Peng, Y.-j., Lilly, S.~J., Kova{\v c}, K., et al.\ 2010, \apj, 721, 193 
\bibitem[Polletta et al.(2007)]{2007ApJ...663...81P} Polletta, M., Tajer, M., Maraschi, L., et al.\ 2007, \apj, 663, 81 
\bibitem[Quintero et al.(2004)]{2004ApJ...602..190Q} Quintero, A.~D., Hogg, D.~W., Blanton, M.~R., et al.\ 2004, \apj, 602, 190 
\bibitem[Scarlata et al.(2007)]{2007ApJS..172..406S} Scarlata, C., Carollo, C.~M., Lilly, S., et al.\ 2007, \apjs, 172, 406  
\bibitem[Schechter \& Press(1976)]{1976ApJ...203..557S} Schechter, P., \& Press, W.~H.\ 1976, \apj, 203, 557 
\bibitem[Scoville et al.(2007)]{2007ApJS..172....1S} Scoville, N., Aussel, H., Brusa, M., et al.\ 2007, \apjs, 172, 1
\bibitem[Snyder et al.(2011)]{2011ApJ...741...77S} Snyder, G.~F., Cox, T.~J., Hayward, C.~C., Hernquist, L., \& Jonsson, P.\ 2011, \apj, 741, 77 
\bibitem[Vulcani et al.(2015)]{2015ApJ...798...52V} Vulcani, B., Poggianti, B.~M., Fritz, J., et al.\ 2015, \apj, 798, 52 
\bibitem[Whitaker et al.(2011)]{whitaker11} Whitaker, K.~E., Labb{\'e}, I., van Dokkum, P.~G., et al.\ 2011, \apj, 735, 86 \bibitem[Whitaker et al.(2012)]{2012ApJ...745..179W} Whitaker, K.~E., Kriek, M., van Dokkum, P.~G., et al.\ 2012, \apj, 745, 179 
\bibitem[Wild et al.(2009)]{2009MNRAS.395..144W} Wild, V., Walcher, C.~J., Johansson, P.~H., et al.\ 2009, \mnras, 395, 144 
\bibitem[Wild et al.(2016)]{2016MNRAS.463..832W} Wild, V., Almaini, O., Dunlop, J., et al.\ 2016, \mnras, 463, 832 
\bibitem[Wong et al.(2012)]{2012MNRAS.420.1684W} Wong, O.~I., Schawinski, K., Kaviraj, S., et al.\ 2012, \mnras, 420, 1684 
\bibitem[Yang et al.(2008)]{2008ApJ...688..945Y} Yang, Y., Zabludoff, A.~I., Zaritsky, D., \& Mihos, J.~C.\ 2008, \apj, 688, 945-971 
\bibitem[Zabludoff et al.(1996)]{1996ApJ...466..104Z} Zabludoff, A.~I., Zaritsky, D., Lin, H., et al.\ 1996, \apj, 466, 104 

\end{thebibliography}
\end{document}